\documentclass[12pt,letterpaper]{JHEP3}
\usepackage{epsfig,amssymb}

\title{A Non-Singular One-Loop Wave Function of
the Universe From a New Eigenvalue 
Asymptotics in Quantum Gravity}

\author{Giampiero Esposito\\
INFN, Sezione di Napoli, and Dipartimento di Scienze
Fisiche, Universit\`a Federico II, Complesso Universitario di
Monte S. Angelo, Via Cintia, Edificio N', 80126 Napoli, Italy\\
E-mail: \email{giampiero.esposito@na.infn.it}}

\author{Guglielmo Fucci\\
Department of Physics, New Mexico Institute of Mining and Technology,
Leroy Place 801, Socorro, NM 87801, USA\\
E-mail: \email{gfucci@nmt.edu}}

\author{Alexander Yu. Kamenshchik\\
Dipartimento di Fisica and INFN,
Sezione di Bologna, Via Irnerio 46, 40126 Bologna, Italy\\
L.D. Landau Institute for Theoretical Physics, Kosygin str. 2,
119334 Moscow, Russia\\
E-mail: \email{alexander.kamenshchik@bo.infn.it}}
 
\author{Klaus Kirsten\\
Department of Mathematics, Baylor University, 
Waco, TX 76798, USA\\
E-mail: \email{Klaus\_Kirsten@baylor.edu}} 

\abstract{Recent work on Euclidean quantum gravity on the 
four-ball has proved regularity at the origin of the generalized
$\zeta$-function built from eigenvalues for metric and ghost modes,
when diffeomorphism-invariant boundary conditions are imposed in 
the de Donder gauge. The hardest part of the analysis involves one
of the four sectors for scalar-type perturbations, the eigenvalues
of which are obtained by squaring up roots of a linear combination 
of Bessel functions of integer adjacent orders, with a coefficient of
linear combination depending on the unknown roots. This paper
obtains, first, approximate analytic formulae for such roots for all values
of the order of Bessel functions. For this purpose, both the descending
series for Bessel functions and their uniform asymptotic expansion
at large order are used.
The resulting generalized $\zeta$-function is also
built, and another check of regularity at the origin
is obtained. For the first time in the literature on quantum gravity
on manifolds with boundary, a vanishing one-loop wave function 
of the Universe is found
in the limit of small three-geometry, which suggests a quantum avoidance
of the cosmological singularity driven by full diffeomorphism invariance 
of the boundary-value problem for one-loop quantum theory.} 

\keywords{Quantum Gravity, Spectral Asymptotics, Zeta Function}

\begin{document}

\section{Introduction}

The subject of boundary effects in quantum field theory 
\cite{Deut79} has always 
received a careful consideration in the literature by virtue of very 
important physical and mathematical motivations, that can be 
summarized as follows.
\vskip 0.3cm
\noindent
(i) Boundary data play a crucial role in the functional-integral
approach \cite{DeWitt03}, 
in the quantum theory of the early universe \cite{Hart83},
in supergravity \cite{Hawk83} and even in string theory \cite{Abou87}.
\vskip 0.3cm
\noindent
(ii) The way in which quantum fields react to the presence of boundaries
is responsible for remarkable physical effects, e.g. the attractive
Casimir force among perfectly conducting parallel plates, which can be
viewed as arising from differences of zero-point energies of the
quantized electromagnetic field \cite{Bord01}.
\vskip 0.3cm
\noindent
(iii) The spectral geometry of a Riemannian manifold 
\cite{Gilk75} with boundary 
is a fascinating problem where many new results have been derived over
the last few years \cite{Kirs01}, \cite{Vass03}.
\vskip 0.3cm
\noindent
(iv) Boundary terms \cite{Moss89} 
in heat-kernel expansions \cite{Avra00} have become a
major subject of investigation in quantum gravity \cite{Vass03},
since they shed new light on one-loop conformal anomalies 
(\cite{Espo94}, \cite{Tsou05})
and one-loop divergences \cite{Espo05a}.

Within such a framework,
recent work by the authors \cite{Espo05} has studied in detail the spectral
asymptotics of Euclidean quantum gravity on the four-ball, motivated
by the longstanding problem of finding local and diff-invariant
boundary conditions on metric perturbations that are compatible with
self-adjointness and strong ellipticity 
(see appendix) of the boundary-value
problem relevant for one-loop quantum theory 
\cite{Espo97}. Interestingly, we have
found that only one of the four eigenvalue conditions (i.e. the 
equations obeyed by the eigenvalues by virtue of boundary
conditions) for scalar modes
is responsible for lack of strong ellipticity that was expected, on
general ground, from the work in \cite{Avra99}. Moreover, the spectral
coefficients $K_{a}^{(j)}$ (see Sect. 5 of \cite{Espo05}) 
occurring in the uniform asymptotic expansion of the logarithmic
derivative of such an eigenvalue condition: 
\begin{equation}
F_{B}^{+}(n,x) \equiv J_{n}'(x)+\left({n\over x}-{x\over 2}\right)
J_{n}(x)=0, \; \forall n \geq 3,
\label{(1.1)}
\end{equation}
ensure the validity, $\forall j=1,{\ldots} ,\infty$, of the peculiar
spectral identity
\begin{equation}
\sum_{a=j}^{4j}{\Gamma(a+1)\over \Gamma(a-j+1)}K_{a}^{(j)}=0.
\label{(1.2)}
\end{equation}
Equation (1.2) engenders, in turn, regularity at the origin of the 
generalized $\zeta$-function built from the eigenvalues
$E_{i}=x_{i}^{2}$, $x_{i}$ being the roots of (1.1) (see below). Thus,
a non-trivial example is found where the $\zeta(0)$ value remains
meaningful even though strong ellipticity of the boundary-value
problem is violated.

The aim of the present paper is to perform a numerical and
analytical investigation of (1.1) which, besides being of intrinsic 
mathematical interest, may be of some help in understanding Euclidean
quantum gravity on the four-ball. Such a background is not an
oversimplification, since simple supergravity with massless
gravitinos, which already makes sense only on Ricci-flat four-manifolds
\cite{Dese76} if the boundary is empty, 
is further restricted to flat Euclidean
four-manifolds if a local description of spin ${3\over 2}$ in terms of
two sets of potentials is exploited 
(see a detailed proof in \cite{Espo96}).

Sections 2 and 3 are devoted to numerical and analytical evaluation
of roots of (1.1). Section 4  
obtains the resulting asymptotic expansion
of the generalized $\zeta$-function, while concluding remarks,
with emphasis on quantum cosmological implications, are
presented in Sect. 5.

\section{Numerical evaluation of roots}

If the Bessel function $J_{n}$ is an even (resp. odd) function of
$x$, its first derivative $J_{n}'$ is an odd (resp. even) function of
$x$. Thus, roots of (1.1) occur always in equal and opposite pairs
$y_{i}=(x_{i},-x_{i})$, and the roots $x_{i}$ are sufficient to build
the generalized $\zeta$-function from the eigenvalues 
$E_{i}=x_{i}^{2}$ \cite{Espo05}. Strictly, we should write 
$E_{i}={x_{i}^{2}\over q^{2}}$, where $q$ is the three-sphere radius,
$S^{3}$ being the boundary in our case, but we set $q=1$ for simplicity.
In \cite{Espo05} we have also studied the eigenvalue condition
\begin{equation}
F_{B}^{-}(n,x) \equiv J_{n}'(x)-\left({n\over x}+{x\over 2}\right)
J_{n}(x)=0, \; \forall n \geq 3,
\label{(2.1)}
\end{equation}
and hereafter we present a table of numerical roots in the two cases.
The vanishing root does not contribute to the
generalized $\zeta$-function, and hence it is not included.

\begin{table}
\caption{The first $5$ roots of $F_{B}^{+}$ if
$n=3,5,20,50,100$.} 
\begin{tabular} {|c|c|c|c|c|c|}
\hline
$F_{B}^{+}$ & 1 & 2 & 3 & 4 & 5
\\
\hline
n=3 & 3.05424 & 6.70613 & 9.96947 & 13.1704 & 16.3475
\\
\hline
n=5 & 4.10467 & 9.0174 & 12.5069 & 15.8305 & 19.0872
\\
\hline
n=20 & 8.72974 & 25.5005 & 30.0311 & 34.0494 & 37.8272
\\
\hline
n=50 & 14.0029 & 57.153 & 62.8403 & 67.7276 & 72.2186
\\
\hline
n=100 & 19.9008 & 108.855 & 115.757 & 121.592 & 126.887
\\
\hline
\end{tabular}
\end{table}

\begin{table}
\caption{The first $5$ roots of $F_{B}^{-}$ if
$n=3,5,20,50,100$.} 
\begin{tabular} {|c|c|c|c|c|c|}
\hline $F_{B}^{-}$ & 1 & 2 & 3 & 4 & 5
\\
\hline n=3 & 6.63569 & 9.94638 & 13.1603 & 16.3422 & 19.5097
\\
\hline n=5 & 8.96599 & 12.4873 & 15.8207 & 19.0815 & 22.3053
\\
\hline n=20 & 25.4909 & 30.0253 & 34.0454 & 37.8242 & 41.4602
\\
\hline n=50 & 57.1508 & 62.8387 & 67.7263 & 72.2175 & 76.4628
\\
\hline n=100 & 108.854 & 115.756 & 121.592 & 126.886 & 131.839
\\
\hline
\end{tabular}
\end{table}

As one can see, there exist a countable infinity of roots 
$y_{i}(F_{B}^{\pm})$ for each value of $n \geq 3$; they are very
close in that
\begin{equation}
\lim_{i \to \infty} \Bigr | y_{i+1}(F_{B}^{+})
-y_{i}(F_{B}^{-}) \Bigr |=0.
\label{(2.2)}
\end{equation}
We also plot, to provide an example,  
the functions $F_{B}^{+}(5,x)$ and $F_{B}^{-}(5,x)$ (for larger
values of $n$, the plots of $F_{B}^{\pm}(n,x)$ are harder to
visualize, since the first root moves away from the origin).

\EPSFIGURE[htp]{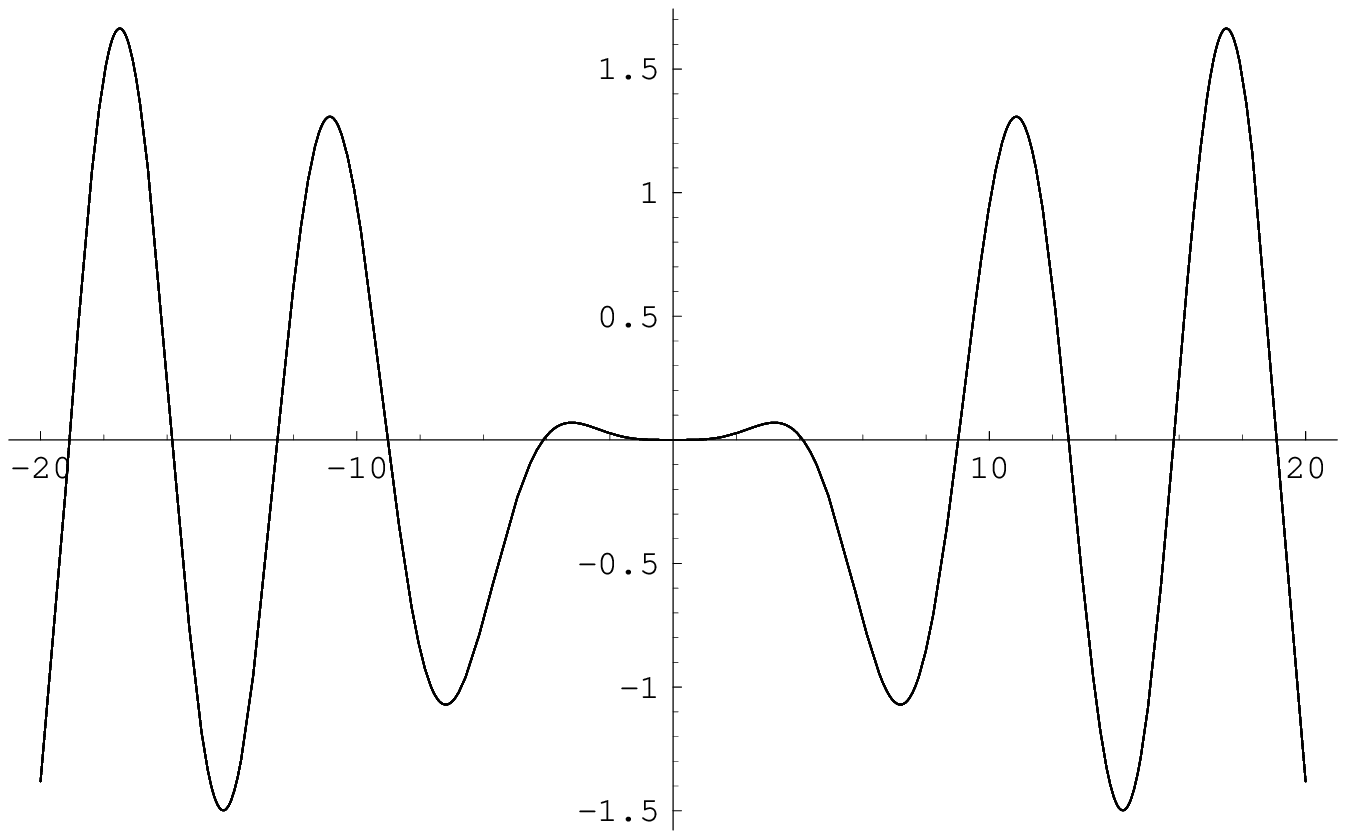, width=.6\textwidth}{The function $F_{B}^{+}(5,x)$.}

\EPSFIGURE[htp]{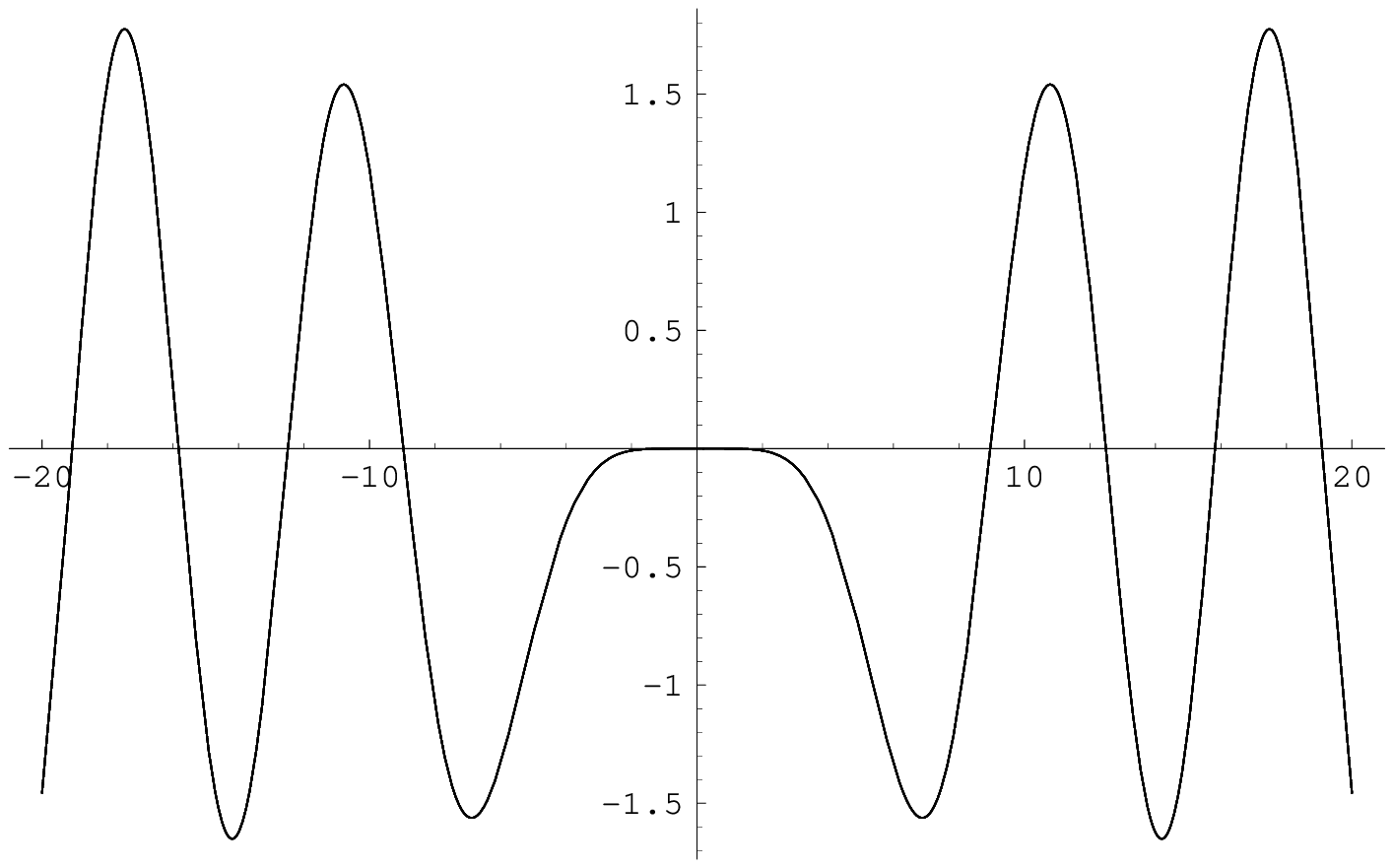, width=.6\textwidth}{The function $F_{B}^{-}(5,x)$.}

\clearpage

\section{Approximate analytic formulae for the roots}

In the course of investigating the zeros of $J_{n}(x)$, or
$J_{n}'(x)$ or suitable combinations of the two, the starting point,
following the seminal work by McMahon \cite{Maho95}, is the 
descending-series formula of Hankel, according to which, for all $n$,
\begin{equation}
J_{n}(x)=\sqrt{2\over \pi x}\left[\varphi_{n}(x)
\cos \left(x-{\pi \over 4}-n{\pi \over 2}\right)
+\psi_{n}(x)\sin \left(x-{\pi \over 4}
-n {\pi \over 2}\right)\right],
\label{(3.1)}
\end{equation}
where
\begin{equation}
\varphi_{n}(x)=1+\sum_{k=1}^{\infty}
{(-1)^{k}\over (2k)! (2x)^{2k}}
{\Gamma \left(n+2k+{1\over 2}\right)\over
\Gamma \left(n-2k+{1\over 2}\right)},
\label{(3.2)}
\end{equation}
\begin{equation}
\psi_{n}(x)=\sum_{k=0}^{\infty}
{(-1)^{k+1}\over (2k+1)! (2x)^{2k+1}}
{\Gamma \left(n+2k+{3\over 2}\right) \over
\Gamma \left(n-2k-{1\over 2}\right)}.
\label{(3.3)}
\end{equation}
We now exploit the identity
\begin{equation}
J_{n}'(x)+{n\over x}J_{n}(x)=J_{n-1}(x)
\label{(3.4)}
\end{equation}
to re-express the eigenvalue condition (1.1) in the form
\begin{equation}
J_{n}(x)-{2\over x}J_{n-1}(x)=0.
\label{(3.5)}
\end{equation}
As a second step, we define
\begin{equation}
\theta_{n}(x) \equiv x-{\pi \over 4}-n{\pi \over 2},
\label{(3.6)}
\end{equation}
and rely upon Eqs. (3.1)--(3.3) to further re-express the eigenvalue 
condition (3.5) in the form (hereafter $m(n) \equiv 4n^{2}$ as
in \cite{Maho95})
\begin{eqnarray}
\tan \theta_{n}(x)&=& {{2\over x}\psi_{n-1}(x)-\varphi_{n}(x) \over 
\psi_{n}(x)+{2\over x}\varphi_{n-1}(x)} \nonumber \\
&=& {8x \over (m-17)}
+\sum_{r=1}^{3}{A_{r}(m(n))\over x^{2r-1}}
+{\rm O}(x^{-7}),
\label{(3.7)}
\end{eqnarray}
where the term linear in $x$ distinguishes $\tan \theta_{n}$ from the
purely Bessel case (in which $J_{n}(x)=0$ and $\tan \theta_{n}$ only has
inverse odd powers of $x$), and we find 
\begin{equation}
A_{r}(m)=(m-17)^{-r-1}\sum_{k=0}^{6r}\alpha_{r,k}m^{k/2},
\label{(3.8)}
\end{equation}
where the only vanishing coefficients up to $x^{-5}$ in (3.7) are
$$
\alpha_{1,3}, \alpha_{1,5}, \alpha_{3,9}, \alpha_{3,11},
\alpha_{5,15}, \alpha_{5,17}.
$$
The first term on the second line of (3.7) is responsible
for the new features with respect to the analysis 
of \cite{Maho95}, where 
the roots of $J_{n}(x)$ {\it for all} $n$ were evaluated approximately 
for the first time. For that problem, the author exploited the fact
that, for an equation of the form
\begin{equation}
x=\beta+{P\over x}+{Q\over x^{3}}+{R\over x^{5}}
+{\rm O}(x^{-7}),
\label{(3.9)}
\end{equation}
the method of iterated approximations gives
\begin{equation}
x=\beta \left[1+{P\over \beta^{2}}+{(Q-P^{2})\over \beta^{4}}
+{(R-4PQ+2P^{3})\over \beta^{6}}+{\rm O}(\beta^{-8})\right].
\label{(3.10)}
\end{equation}
In our case we have to find, from (3.7), approximate solutions
of the equation ($s$ being any integer, here introduced to take
into account periodicity of the $\tan$ function)
\begin{equation}
\theta_{n}(x)-s\pi ={\rm arctan} 
\left({8x \over (m-17)}
+\sum_{r=1}^{3}
{A_{r}(m(n))\over x^{2r-1}}+{\rm O}(x^{-7})\right).
\label{(3.11)}
\end{equation}
On defining the variable $w \equiv x^{-1}$, the right-hand side of  
(3.11) can be Taylor expanded about $w=0$, and this leads to the equation
\begin{equation}
x=\beta(s,n)+\sum_{r=1}^{3}{B_{r}(m(n))\over x^{2r-1}}
+{\rm O}(x^{-7}),
\label{(3.12)}
\end{equation}
which is of the form (3.9), with (here $s \geq 0$)
\begin{equation}
\beta(s,n) \equiv \pi \left(s+{n\over 2}+{3\over 4}\right),
\label{(3.13)}
\end{equation}
\begin{equation}
B_{1}(m)=P= -{(m-17)\over 8},
\label{(3.14)}
\end{equation}
\begin{equation}
B_{2}(m)=Q=-{1721\over 384}+2m^{1/2}-{35\over 192}m
-{1\over 384}m^{2}, 
\label{(3.15)}
\end{equation}
\begin{equation}
B_{3}(m)=R={79201\over 5120}-{47\over 4}m^{1/2}
+{14973 \over 5120}m-{1 \over 4}m^{3/2} 
+ {7m^{2}\over 1024}-{m^{3}\over 5120}.
\label{(3.16)}
\end{equation}
Our approximate roots read therefore as
\begin{equation}
x(s,n) \sim \beta(s,n)\left[1+{\gamma_{1}\over \beta^{2}(s,n)}
+{\gamma_{2} \over \beta^{4}(s,n)}
+{\gamma_{3} \over \beta^{6}(s,n)}
+{\rm O}(\beta^{-8})\right],
\label{(3.17)}
\end{equation}
where, from (3.10),
\begin{equation}
\gamma_{1}=B_{1}=-{(m-17)\over 8}, 
\label{(3.18)}
\end{equation}
\begin{equation}
\gamma_{2}=B_{2}-B_{1}^{2}
=-{3455\over 384}+2m^{1/2}+{67\over 192}m-{7\over 384}m^{2},
\label{(3.19)}
\end{equation}
\begin{eqnarray}
\gamma_{3}&=& B_{3}-4B_{1}B_{2}+2B_{1}^{3} \nonumber \\
&=&{1117523\over 15360}-{115\over 4}m^{1/2}-{5907\over 5120}m
+{3\over 4}m^{3/2}+{421\over 3072}m^{2}-{83\over 15360}m^{3},
\label{(3.20)}
\end{eqnarray}
which provides a good approximation at large $x$ and 
low values of $n$, as is further discussed below.
Unlike the Dirichlet or Neumann problems, our $\gamma_{p}$ are,
in general, non-analytic functions of $m$.

Note that large $x$ and small $n$ actually also means large $s$;
this is the only way Eq. (3.17) provides a systematic asymptotic 
expansion with smaller and smaller correction terms.
The roots in Table 3 are well approximated by their analytic 
form (3.17) at large $x$ and low $n$.
For $n=3,5$, the first two lines display all numerical roots of 
(3.5) from the first to the fifth, while the third and fourth
line show their approximation $x(s,3)$ and $x(s,5)$ from
(3.17), with $s=0,1,2,3,4$ (recall that the first root is obtained
for $s=0$). Remarkably, Eq. (3.17)--(3.20) approximate very well all roots
starting from $x(1,n)$, while $x(0,n)$ is fairly well approximated
only for the lowest value of $n$, i.e. $n=3$.

\begin{table}
\caption{Numerical roots of (3.5).}
\begin{tabular} {|c|c|c|c|c|c|}
\hline
$F_{B}^{+}$ & 1 & 2 & 3 & 4 & 5 
\\
\hline
n=3 & 3.05424 & 6.70613 & 9.96947 & 13.1704 & 16.3475 
\\
\hline
n=5 & 4.10467 & 9.0174 & 12.5069 & 15.8305 & 19.0872 
\\
\hline
x(s,3) & 3.13152 & 6.70672 & 9.96951 & 13.1704 & 16.3475 
\\
\hline
x(s,5) & 5.00988 & 9.03333 & 12.5089 & 15.831 & 19.0873 
\\
\hline
\end{tabular}
\end{table}

The expansion (3.17) of the roots of (3.5), 
which holds for small $n$ and large $x$, should be replaced by a
different expansion at large $n$. For this purpose, one may start 
from the uniform asymptotic expansion
of $J_{n}(x)$ and $J_{n}'(x)$ \cite{Olve54}. In terms
of $x$ the desired formulae read as (our polynomials $u_{k}$ and
$v_{k}$ below can be found, for example, in 
appendix B of \cite{Espo05})
\begin{equation}
J_{n}(x) \sim {e^{n {\tilde \eta}(n,x)} \over \sqrt{2\pi n}
\left(1-{x^{2} \over n^{2}}\right)^{1 \over 4}}
\left \{ 1+\sum_{k=1}^{\infty}{u_{k}(t(n,x)) 
\over n^{k}} \right \},
\label{(3.21)}
\end{equation}
and
\begin{equation}
J_{n}'(x) \sim \sqrt{n \over 2\pi} \; {1 \over x}
\left(1-{x^{2}\over n^{2}}\right)^{1 \over 4}e^{n{\tilde \eta}(n,x)}
\left \{ 1+\sum_{k=1}^{\infty}{v_{k}(t(n,x)) 
\over n^{k}} \right \},
\label{(3.22)}
\end{equation}
where
\begin{equation}
{\tilde \eta}(n,x) \equiv {1 \over n}\sqrt{n^{2}-x^{2}}-\log
\left[{n \over x}+\sqrt{{n^{2} \over x^{2}}-1}\right],
\label{(3.23)}
\end{equation}
and
\begin{equation}
t(n,x) \equiv {1 \over \sqrt{1-{x^{2} \over n^{2}}}}.
\label{(3.24)}
\end{equation}
On inserting (3.21) and (3.22) in (1.1), we obtain
\begin{equation}
F_{B}^{+} \sim {e^{n {\tilde \eta}}\over \sqrt{2\pi n}}
{(1+t)\over \sqrt{t}}\left(1-t^{-2}\right)^{-1/2} 
\left[1+{(1-t)\over 2t}(n+u_{1})
+\sum_{k=1}^{\infty}{g_{k}(t)\over n^{k}}\right],
\label{(3.25)}
\end{equation}
where 
\begin{equation}
g_{k}(t) \equiv {v_{k}(t)+t u_{k}(t) \over (1+t)}
+{(1-t)\over 2t}u_{k+1}(t).
\label{(3.26)}
\end{equation}
It is very important, at this stage, to note explicitly that the
expansion of our equation written above only holds for $x<n$. This
can be understood on considering the functional form of (3.25)
where, from (3.21) and (3.22), the 
variable $x$ satisfies the inequality $x<n$. For this reason
we can only find approximate solutions at large $n$
of (3.5) when $x$ is smaller
than $n$. Looking at the table of numerical roots of our
equation one can clearly note that, for large $n$, there is always
only one root satisfying the condition $x<n$. Then the expansion
here described can be used for approximating such roots.

For this purpose, let us define the following function (cf. (3.25)):
\begin{equation}
y_{\lambda}(n,x) \equiv 1+{(1-t)\over 2t}(n+u_{1})
+\sum_{k=1}^{\lambda}{g_{k}(t)\over n^{k}}.
\label{(3.27)}
\end{equation}
One can obtain a good approximation of the roots by considering
only the first three terms of the series which appears in the last
equation. In this way one can get approximate values 
for the roots from the zeros of $y_{\lambda}(n,x)$ 
which are summarized, for a few values of $n$, in Table 4.
An approximate formula for the roots $x(n)$ is obtained by
setting $y_{\lambda}(n,x)=0$. To leading order we find that such an
equation reduces to
$$
{\sqrt{n^{2}-x^{2}}\over x}+{n\over x}-{x\over 2}=0,
$$
which is solved by $x=\pm 2\sqrt{n}\sqrt{1-{1\over n}}$. We therefore write
$x=2\sqrt{n}+c$, with $c$ a constant. If we plug it into Eq. (3.27) and 
expand for large values of $n$, and choose $c$ so that the leading order
vanishes, we find $c=0$. Then we continue with $x \sim 2\sqrt{n}+0
+{c\over \sqrt{n}}$ and proceed in the same way to find $c=-1$. In this 
way we obtain, eventually,
\begin{equation}
x \sim \pm \left[2\sqrt{n}-{1\over \sqrt{n}}+{3\over 4}n^{-3/2}
+{11\over 8}n^{-5/2}-{157\over 64}n^{-7/2}
+{\rm O}(n^{-9/2})\right].
\label{(3.28)}
\end{equation}
By looking at the first two terms of this expansion we discover that a
better expansion parameter is actually $(n-1)$, and hence we obtain
\begin{equation}
x \sim \pm 2\sqrt{n-1}\left[1+{1\over 2}(n-1)^{-2}
-{17\over 8}(n-1)^{-4}+{\rm O}((n-1)^{-6})\right].
\label{(3.29)}
\end{equation}
The fact that only every second term is non-vanishing
in (3.29) suggests this is a good expansion parameter. It also explains why it
approximates better than the expansion in terms of $n$. As the leading 
two terms show, an expansion of $x$ in terms of $n-\delta$ for some
$\delta \in ]0,1[$ is not possible, because for large values of $n$
one cannot satisfy the equation $y_{\lambda}(n,x)=0$ from Eq. (3.27).

\begin{table}
\caption{Exact vs. approximate roots of (3.5) 
at large $n$, with $x<n$.}
\begin{tabular} {|c|c|c|c|c|}
\hline $F_{B}^{+}$ & $n=20$ & $n=50$ & $n=70$ & $n=100$
\\
\hline $x(n)$ & 8.72974 & 14.0029 & 16.6152 & 19.9008
\\
\hline $y_{3}(n,x)$ & 8.9376 & 14.0029 & 16.6149 & 19.9007
\\
\hline
\end{tabular}
\end{table}

Oscillating asymptotic expansions of the roots as $x >n$ can also be
obtained, but they are not written down for brevity. The leading term
(3.13) in the expansion (3.12) of the roots can be used to reproduce 
two of the three contributions to $\zeta(0)$ found in \cite{Espo05}.
This is reassuring, but there is no exact formula for $x(s,n)$ for all
$s,n$, and hence it remains unclear how to evaluate $\zeta(0)$ without
resorting to contour integration (see below).

\section{Generalized $\zeta$-functions from contour integrals}

Tables 1 and 2 jointly with the limit in (2.2) tell us that, for
any fixed $n$, the root $x(i,n,+)$ of $F_{B}^{+}$ is very close to
and slightly larger than the root $x(i-1,n,-)$ of $F_{B}^{-}$. Thus,
a $\rho(i,n)$ positive and much smaller than $1$ exists such that,
for the eigenvalues obtained by squaring up the roots, one can write
\begin{equation}
E(i,n,+)=E(1,n,+)\delta_{i,1}+E(i-1,n,-)(1+\rho(i,n))
(1-\delta_{i,1}),
\label{(4.1)}
\end{equation}
for all $n \geq 3$ and for all $i \geq 1$, where $i$ labels here the
countable infinity of roots for any given $n$ and hence starts
from $1$. 

This suggests looking for a link among the generalized $\zeta$-functions
$\zeta_{B}^{+}(s)$ and $\zeta_{B}^{-}(s)$ (hereafter $s$ is the independent
variable in the generalized $\zeta$-functions). 
Following the methods developed in \cite{Bord96a, Bord96b},
our starting point is the integral representation in 
\cite{Espo05}, i.e. (hereafter
$\beta_{+} \equiv n, \beta_{-} \equiv n+2$)
\begin{eqnarray}
\zeta_{B}^{\pm}(s)& \equiv &{(\sin \pi s)\over \pi}\sum_{n=3}^{\infty}
n^{-(2s-2)} \nonumber \\
& \; & \int_{0}^{\infty}
dz \; z^{-2s}{\partial \over \partial z}{\rm log}
\left[z^{-\beta_{\pm}(n)}\left(znI_{n}'(zn)+\left({z^{2}n^{2}\over 2}
\pm n \right)I_{n}(zn)\right)\right].
\label{(4.2)}
\end{eqnarray}
We exploit the uniform asymptotic expansion of modified
Bessel functions and their first derivatives to find (hereafter
$\tau \equiv (1+z^{2})^{-1/2}$) 
\begin{equation}
znI_{n}'(zn)+\left({z^{2}n^{2}\over 2} \pm n \right)I_{n}(zn)
\sim {n^{2}\over 2 \sqrt{2 \pi n}}{e^{n \eta} \over \sqrt{\tau}}
\left({1\over \tau}-\tau \right) 
\left (1+ \sum_{k=1}^{\infty}{r_{k,\pm}(\tau)\over n^{k}}
\right),
\label{(4.3)}
\end{equation}
where we have (bearing in mind that $u_{0}=v_{0}=1$)
\begin{equation}
r_{k,\pm}(\tau) \equiv u_{k}(\tau)
+{2\tau \over (1-\tau^{2})}
\Bigr((v_{k-1}(\tau) \pm \tau 
u_{k-1}(\tau) \Bigr),
\label{(4.4)}
\end{equation}
for all $k \geq 1$, with $u_{k}$ and $v_{k}$ as in (3.21)
and (3.22). Hereafter we set
\begin{equation}
\Omega \equiv \sum_{k=1}^{\infty}{r_{k,\pm}(\tau(z))\over n^{k}},
\label{(4.5)}
\end{equation}
and rely upon the formula
\begin{equation}
\log(1+\Omega) \sim \sum_{k=1}^{\infty}(-1)^{k+1}
{\Omega^{k}\over k}
\label{(4.6)}
\end{equation}
to evaluate the uniform asymptotic expansion 
\begin{equation}
{\rm log}\left(1+\sum_{k=1}^{\infty}{r_{k,\pm}(\tau(z))\over n^{k}}
\right) \sim \sum_{k=1}^{\infty}{R_{k,\pm}(\tau(z))\over n^{k}}.
\label{(4.7)}
\end{equation}
Hence we find \cite{Espo05}
\begin{equation}
R_{1,\pm}=(1 \mp \tau)^{-1}
\left({17\over 8}\tau \mp{1\over 8}\tau^{2}-{5\over 24}\tau^{3}
\pm{5\over 24}\tau^{4}\right),
\label{(4.8)}
\end{equation}
\begin{equation}
R_{2,\pm}=(1 \mp \tau)^{-2}
\left(-{47\over 16}\tau^{2} \pm{15\over 8}\tau^{3}-{21\over 16}\tau^{4}
\pm{3\over 4}\tau^{5}-{1\over 16}\tau^{6} \mp{5\over 8}\tau^{7}
+{5\over 16}\tau^{8}\right),
\label{(4.9)}
\end{equation}
\begin{eqnarray}
R_{3,\pm}&=& (1 \mp \tau)^{-3}\biggr(
{1721\over 384}\tau^{3} \mp{441\over 128}\tau^{4}+{597\over 320}\tau^{5}
\mp{1033\over 960}\tau^{6} 
+{239\over 80}\tau^{7} \nonumber \\
&\mp & {28\over 5}\tau^{8}
+{2431\over 576}\tau^{9} \pm{221\over 192}\tau^{10} 
- {1105\over 384}\tau^{11} \pm{1105\over 1152}\tau^{12}\biggr),
\label{(4.10)}
\end{eqnarray}
and, in general,
\begin{equation}
R_{j,\pm}(\tau(z))=(1 \mp \tau)^{-j}\sum_{a=j}^{4j}C_{a}^{(j,\pm)}
\tau^{a}.
\label{(4.11)}
\end{equation}
Note that, at $\tau=1$ (i.e. $z=0$), our $r_{k,+}(\tau)$ and $R_{k,+}(\tau)$
are singular. Such a behaviour
is not seen for any of the strongly elliptic boundary-value
problems \cite{Kirs01}. 

The $\zeta_{B}^{-}(s)$ function is more easily dealt with.
It indeed receives contributions from terms equal to 
\begin{eqnarray}
B_{-}(s)& \equiv & \sum_{n=3}^{\infty}n^{-(2s-2)}{(\sin \pi s)\over \pi}
\int_{0}^{\infty}dz \; z^{-2s}{\partial \over \partial z}
\log \left({{1\over \tau(z)}-\tau(z) \over z^{2}}\right)
\nonumber \\
&=& \omega_{0}(s){(\sin \pi s)\over \pi}
\int_{0}^{\infty}dz \; z^{-2s}
{\partial \over \partial z} \log {1\over \sqrt{1+z^{2}}}
=-{1\over 2}\omega_{0}(s),
\label{(4.12)}
\end{eqnarray}
and $\sum_{j=1}^{\infty}B_{j,-}(s)$, having defined, with  
$\lambda=0,j$ 
\begin{equation}
\omega_{\lambda}(s) \equiv \sum_{n=3}^{\infty}n^{-(2s+\lambda-2)}
=\zeta_{H}(2s+\lambda-2;3),
\label{(4.13)}
\end{equation}
\begin{equation}
B_{j,-}(s) \equiv \omega_{j}(s){(\sin \pi s)\over \pi}
\int_{0}^{\infty}dz \; z^{-2s}{\partial \over \partial z}
R_{j,-}(\tau(z)).
\label{(4.14)}
\end{equation}
On using the same method as in \cite{Espo05}, such formulae lead to 
\begin{equation}
\zeta_{B}^{-}(0)=-{5\over 4}+{1079\over 240}+{5\over 2}
-{1\over 16}\sum_{a=3}^{12}C_{a}^{(3,-)} 
={206\over 45},
\label{(4.15)}
\end{equation}
a result which agrees with a derivation of $\zeta_{B}^{-}(0)$ 
relying upon the contour method in \cite{Barv92}.

To deal with the generalized $\zeta$-function $\zeta_{B}^{+}$
we define, in analogy with Eq. (4.12),
\begin{equation}
B_{+}(s)\equiv \omega_{0}(s){(\sin \pi s)\over \pi}
\int_{0}^{\infty}dz \; z^{-2s}
{\partial \over \partial z} \log \left({1\over \tau(z)}
-\tau(z)\right),
\label{(4.16)}
\end{equation}
and, in analogy to Eq. (4.14),
\begin{equation}
B_{j,+}(s) \equiv \omega_{j}(s){(\sin \pi s)\over \pi}
\int_{0}^{\infty}dz \; z^{-2s}{\partial \over \partial z}
R_{j,+}(\tau(z)).
\label{(4.17)}
\end{equation}
Unlike our work in \cite{Espo05} we now exploit Eq. (4.4) to evaluate
\begin{equation}
r_{k,+}(\tau)-r_{k,-}(\tau)={4\tau^{2}\over (1-\tau^{2})}u_{k-1}(\tau),
\label{(4.18)}
\end{equation}
and hence we find
\begin{equation}
R_{1,+}=R_{1,-}+{4\tau^{2}\over (1-\tau^{2})},
\label{(4.19)}
\end{equation}
\begin{equation}
R_{2,+}=R_{2,-}+{4\tau^{2}\over (1-\tau^{2})}
\left(u_{1}-{2\tau^{2}\over (1-\tau^{2})}-R_{1,-}\right),
\label{(4.20)}
\end{equation}
\begin{eqnarray}
R_{3,+}&=& R_{3,-}+{4\tau^{2}\over (1-\tau^{2})}
\left(u_{2}-{4\tau^{2}\over (1-\tau^{2})}u_{1}
-u_{1}R_{1,-}-R_{2,-}
+{4\tau^{2}\over (1-\tau^{2})}R_{1,-}\right) \nonumber \\
&+& {64\over 3}{\tau^{6}\over (1-\tau^{2})^{3}}
+{2\tau^{2}\over (1-\tau^{2})}R_{1,-}^{2},
\label{(4.21)}
\end{eqnarray}
and so on. This makes it possible to evaluate $B_{j,+}(s)-B_{j,-}(s)$
for all $j=1,2,... \infty$. 
Only $j=3$ contributes to $\zeta_{B}^{\pm}(0)$, because
\begin{equation}
\omega_{j}(s){(\sin \pi s)\over \pi} \sim
{1\over 2}\delta_{j,3}+{\tilde b}_{j,1}s+{\rm O}(s^{2}),
\label{(4.22)}
\end{equation}
where \cite{Espo05}
\begin{equation}
{\tilde b}_{j,1}=-1-2^{2-j}+\zeta_{R}(j-2)(1-\delta_{j,3})
+\gamma \delta_{j,3},
\label{(4.23)}
\end{equation}
and we find
\begin{eqnarray}
B_{3,+}(s)&=& B_{3,-}(s) \nonumber \\
&-& \omega_{3}(s){(\sin \pi s)\over \pi}
\lim_{\mu \to 1} \int_{0}^{\mu}d\tau \; \tau^{2s}(1-\tau)^{-s}
(1+\tau)^{-s}{\partial \over \partial \tau}
(R_{3,+}-R_{3,-}).
\label{(4.24)}
\end{eqnarray}
Such a limit as $\mu$ tends to $1$ means that we split the original
integration interval according to \cite{Espo05}
$$
\int_{0}^{1}d\tau=\int_{0}^{\mu}d\tau+\int_{\mu}^{1}d\tau.
$$
In the first interval on the right-hand side we can integrate 
at small $\mu$ the terms involving
negative powers of $(1-\tau)$ in the uniform asymptotics. On the
other hand, since the original integral from $0$ to $1$ is
independent of $\mu$, we can, at last, take the $\mu \to 1$
limit \cite{Espo05}. On defining ${\widetilde \mu} \equiv 
\left({1\over \mu^{2}}-1 \right)^{1/2}$, and writing again $x_{i}$
for the roots of Eq. (1.1), the limit as $\mu \rightarrow 1$ of the
integral $\int_{\mu}^{1}d\tau$ is reducible to a sum of limits 
$$
\lim_{{\widetilde \mu} \to 0} \int_{0}^{{\widetilde \mu}}
{z^{1-2s}\over x_{i}^{2}+z^{2}}dz,
$$
each of which vanishes since such integrals are equal to
$$
{{\widetilde \mu}^{2-2s}
{}_{2}F_{1}(1,1-s,2-s,-{\widetilde \mu}^{2}/x_{i}^{2})
\over x_{i}^{2}(2-2s)}.
$$

The derivative in the integrand on the right-hand side of 
Eq. (4.24) reads as 
\begin{equation}
{\partial \over \partial \tau}(R_{3,+}-R_{3,-})
=(1-\tau)^{-4}(1+\tau)^{-4}\Bigr(80\tau^{3}-24 \tau^{5}
+32 \tau^{7}-8 \tau^{9}\Bigr).
\label{(4.25)}
\end{equation}
We can thus use the definition
\begin{equation}
Q_{\mu}(\alpha,\beta,\gamma) \equiv \int_{0}^{\mu}
\tau^{\alpha}(1-\tau)^{\beta}(1+\tau)^{\gamma} d\tau,
\label{(4.26)}
\end{equation}
where $Q_{\mu}$ can be evaluated from a hypergeometric function
of 2 variables according to \cite{Espo05}
\begin{equation}
Q_{\mu}(\alpha,\beta,\gamma)={\mu^{\alpha+1}\over \alpha+1}
F_{1}(\alpha+1,-\beta,-\gamma,\alpha+2;\mu,-\mu),
\label{(4.27)}
\end{equation}
to express (4.24) through the functions 
$Q_{\mu}(2s+a,-s-4,-s-4)$, with $a=3,5,7,9$. This leads to
\begin{eqnarray}
\zeta_{B}^{+}(0)&=&\zeta_{B}^{-}(0)+B_{3,+}(0)-B_{3,-}(0) \nonumber \\
&=& \zeta_{B}^{-}(0)-{1\over 24}\sum_{l=1}^{4}
{\Gamma(l+1)\over \Gamma(l-2)}\left[\psi(l+2)-{1\over (l+1)}\right]
\kappa_{2l+1}^{(3)} \nonumber \\
&=&{206\over 45}+2={296\over 45},
\label{(4.28)}
\end{eqnarray}
where $\kappa_{2l+1}^{(3)}$ are the four coefficients 
of odd powers of $\tau$ on the right-hand side of (4.25).
Regularity of $\zeta_{B}^{+}(s)$ at the origin is guaranteed because
$\lim_{s \to 0}s \zeta_{B}^{+}(s)$ is proportional to
\begin{equation}
\sum_{l=1}^{4}{\Gamma(l+1)\over \Gamma(l-2)}\kappa_{2l+1}^{(3)}=0,
\label{(4.29)}
\end{equation}
which is a particular case of the peculiar spectral cancellation
(cf. Eq. (1.2)) 
\begin{equation}
\sum_{a=a_{\rm min}(j)}^{a_{\rm max}(j)}
{\Gamma \left({(a+1)\over 2}\right)\over 
\Gamma \left({(a+1)\over 2}-j \right)}
\kappa_{a}^{(j)}=0,
\label{(4.30)}
\end{equation}
where $a$ takes both odd and even values. The case $j=3$ is simpler because
then only $\kappa_{a}^{(j)}$ coefficients with odd $a$ are non-vanishing.

Remaining contributions to $\zeta(0)$, being obtained from strongly elliptic
sectors of the boundary-value problem, are easily found to agree with the
results in \cite{Espo97}, and we find from transverse-traceless, vector,
scalar and ghost modes the full $\zeta(0)$ value
\begin{eqnarray}
\zeta(0)&=& -{278\over 45}+{494\over 45}-{15\over 2}-17
+{146\over 45}+{757\over 90}+{206 \over 45}+{296 \over 45}
\nonumber \\
&-& {149\over 45}+{77\over 90}+{5\over 2}
={142\over 45}.
\label{(4.31)}
\end{eqnarray}
Since the one-loop prefactor scales as the three-sphere radius 
raised to a power equal to $\zeta(0)$ \cite{Schl85}, we 
find therefore a vanishing one-loop wave function of the Universe at small
three-geometry, which suggests an intriguing quantum avoidance of the
cosmological singularity driven by full diffeomorphism invariance.
As far as we know, this positive $\zeta(0)$ value for pure gravity
is new in the literature. It might have been obtained from our earlier
analysis in \cite{Espo05}, but, at that time, the cross-check provided
by our Eqs. (4.18)--(4.30) was missing, and hence we had not yet reached
this conclusion. It has been here our aim to focus on pure gravity,
but scalar, spinor and gauge-field contributions to the one-loop wave function
of the Universe may be included, if necessary \cite{Espo97}. 

\section{Concluding remarks}

The idea that the infinite-dimensional invariance group determines 
completely not only the action functional but also the boundary 
conditions in (quantum) field theory is very appealing, but such
a program faced severe difficulties after the proof in \cite{Avra99}
that diffeomorphism-invariant boundary conditions in the de Donder 
gauge in quantum gravity are incompatible with strong ellipticity.

The more recent work in \cite{Espo05} by the authors changed a lot
the overall perspective: although the global heat-kernel asymptotics
becomes in general ill-defined, it remains possible to define and
evaluate the pure gravity $\zeta(0)$ value at least on the
Euclidean four-ball, since the associated generalized $\zeta$-function
remains regular at the origin. More precisely, the integral representation
(4.2) of the generalized $\zeta$-function is legitimate because, for any
fixed $n$, there is a countable infinity of roots $x_{j}$ and $-x_{j}$
of Eq. (1.1) and they grow approximately linearly with the integer $j$
counting such roots. The function $F_{B}^{\pm}$ admits therefore a
canonical-product representation which ensures that the integral 
representation (4.2) reproduces the standard definition of generalized
$\zeta$-function \cite{Espo05}. Moreover, even though the Mellin transform 
relating $\zeta$-function to integrated heat kernel cannot be exploited
when strong ellipticity is not fulfilled, it remains possible to define a
generalized $\zeta$-function. For this purpose, a weaker assumption 
provides a sufficient condition, i.e. the existence of a sector in the
complex plane free of eigenvalues of the leading symbol 
of the differential operator under consideration \cite{Seel67}.
To make sure we have not overlooked some properties of the spectrum, we
have been looking for negative eigenvalues or zero-modes, but finding none.
Indeed, negative eigenvalues $E$ would imply purely imaginary roots 
$x=iy$ of Eqs. (1.1) and (2.1), but such roots do not exist, as one can
check both numerically and analytically; zero-modes would be non-trivial
eigenfunctions belonging to zero-eigenvalues, but all modes (tensor,
vector, scalar and ghost modes) are combinations of Bessel functions
\cite{Espo05} for which this is impossible. As far as we can see, we
still find sources of singularities at the origin
in the generalized $\zeta$-function resulting from lack of strong 
ellipticity, but the particular symmetries of the Euclidean 4-ball
background reduce them to the four terms in Eq. (4.29), which add up
to zero despite two of them are non-vanishing.

The present paper has provided an important and possibly simpler
cross-check of the results in \cite{Espo05}. 
This is relevant for one-loop quantum
cosmology in the Hartle--Hawking program \cite{Hart83}, since the
wave function of the Universe at small three-geometry 
\cite{Schl85} becomes a wave function about the four-ball 
background \cite{Espo97}. It is an open question whether it can
also be relevant for
the AdS/CFT correspondence by virtue of the profound link between
AdS/CFT and the Hartle--Hawking 
wave function of the Universe stressed in Sect.
4.3 of \cite{Horo03}, the main problem being that AdS/CFT relies upon
boundary conditions for metric and other fields at infinity, unlike
the case of closed quantum cosmologies.

What is more important, if the Universe had a semiclassical origin
\cite{Hawk05}, our one-loop calculations acquire new interest, further
strengthened by the positive $\zeta(0)$ value in Eq. (4.31): a positive
$\zeta(0)$ value expressing a regular and indeed vanishing one-loop
wave function of the Universe in the limit of small three-geometry.
We propose to interpret this property as an (unexpected) indication
that full diffeomorphism invariance of the boundary-value
problem engenders a quantum avoidance of the cosmological singularity. This
appears as a conceptually novel perspective, and of course further 
hard thinking is in order. More precisely, the work by Schleich
\cite{Schl85} had found that, on restricting the functional integral to
transverse-traceless perturbations, the one-loop semiclassical approximation 
to the wave function of the Universe diverges at small volumes, at least
for the geometry of a three-sphere. The divergence of the wave functional 
does not necessarily mean that the probability density of the wave functional
diverges at small volumes, since the probability density $p[h]$ on the
space of wave functionals $\psi[h]$ is given by
\begin{equation}
p[h]=m[h]|\psi|^{2}[h],
\label{(5.1)}
\end{equation}
which includes the effect of the measure $m[h]$ on this space. The scaling
of this measure is not known in general. In our manifestly covariant
evaluation of the one-loop functional integral for the wave function of
the Universe, it would seem untenable to assume that the measure
$m[h]$ scales in such a way as to cancel exactly the contribution of
$|\psi|^{2} \propto (S^{3}-{\rm radius})^{2\zeta(0)}$. Thus, we conclude
that our one-loop wave function of the Universe vanishes at small volume.
The normalizability condition of the wave function in the limit of
small three-geometry, which is weaker than the condition of its vanishing
in this limit, was instead formulated and studied in \cite{Barv90}. 

Other promising applications of our investigation of the generalized
$\zeta$-function deal with the one-loop effective action for braneworld
models, as is extensively discussed in \cite{Barv05}.

\appendix
\section{Strong ellipticity}

Let $A$ be an elliptic differential operator with leading symbol
$A_{m}(x,\xi)$ and let ${\mathcal K}$ be a cone containing $0$ such that,
for $\xi \not =0$, the spectrum of $A_{m}(x,\xi)$ lies in the
complement of ${\mathcal K}$. 
Let $b^{(0)}(y,\omega,D_{r})$ be the leading partial
symbol of the boundary operator ${\mathcal B}$. The boundary-value
problem $(A,{\mathcal B})$ is then said to be strongly elliptic if,
for 
$$
(0,0) \not = (\omega,\lambda) \in {\partial}{\bf R}_{+}^{D}
\times {\mathcal K},
$$
the equations \cite{Kirs01} (hereafter, $r$ is the geodesic
distance to the boundary)
\begin{equation}
A_{m}(y,0,\omega,D_{r})f(r)=\lambda f(r),
\label{(A.1)}
\end{equation}
\begin{equation}
\lim_{r \to \infty}f(r)=0,
\label{(A.2)}
\end{equation}
\begin{equation}
b^{(0)}(y,\omega,D_{r})f(r)[r=0]=g(\omega),
\label{(A.3)}
\end{equation}
have a unique solution.

For example, in the case of Dirichlet boundary conditions
\begin{equation}
\Bigr[{\mathcal B}\phi \Bigr]_{\partial M}
=[\phi]_{\partial M}=0,
\label{(A.4)}
\end{equation}
with $A$ an operator of Laplace type, one has the leading symbol
$A_{2}(x,\xi)=|\xi|^{2}$, and the general equation (A.1) becomes
the differential equation
\begin{equation}
A_{2}(y,0,\omega,D_{r})f(r)=\left(-{d^{2}\over dr^{2}}
+|\omega|^{2}\right)f(r)=\lambda f(r).
\label{(A.5)}
\end{equation}
The general solution of Eq. (A.5) reads as
\begin{equation}
f(r)=\alpha(\omega)e^{-r \Lambda}+\beta(\omega)e^{r \Lambda},
\label{(A.6)}
\end{equation}
having defined $\Lambda \equiv \sqrt{|\omega|^{2}-\lambda}$. The
asymptotic condition (A.2) for $r \rightarrow \infty$ leads to
$\beta=0$, while Eq. (A.3) engenders, by virtue of Eq. (A.4),
$\alpha(\omega)=g(\omega)$, since the leading partial symbol of the
boundary operator reduces to the identity in the Dirichlet case.
The boundary conditions (A.4) are therefore strongly elliptic with
respect to the cone ${\bf C}-{\bf R}_{+}$.

\acknowledgments
The work of G. Esposito has been partially
supported by PRIN {\it SINTESI}. K. Kirsten is grateful to the Baylor
University Research Committee, to the Max-Planck-Institute for Mathematics
in the Sciences (Leipzig, Germany) and to the INFN for financial support.
The work of A.Yu. Kamenshchik 
was partially supported by the Russian Foundation for
Basic Research under the Grant No. 02-02-16817 and by the Scientific School
Grant No. 2338.2003.2

\end{document}